\documentclass[aps,prl,twocolumn]{revtex4}
\usepackage{amsmath,bm,epsfig}

\def\Fbox#1{\vskip1ex\hbox to 8.5cm{\hfil\fboxsep0.3cm\fbox{%
  \parbox{8.0cm}{#1}}\hfil}\vskip1ex\noindent}  

\newcommand{\sFrac}[2]{{\textstyle\frac{#1}{#2}}}
\let \= \equiv \let\*\cdot \let\~\widetilde \let\-\overline

\begin{document}
\title{What Determines the Yield Stress in Amorphous Solids?}
\author{Smarajit Karmakar, Edan Lerner, Itamar Procaccia and Jacques Zylberg}
\affiliation{Department of Chemical Physics, The Weizmann
 Institute of Science, Rehovot 76100, Israel }
\date{\today}
\begin{abstract}
A crucially important material parameter for all amorphous solids is the yield stress, which is the value of the stress for
which the material yields to plastic flow when it is strained quasi-statically at zero temperature. It is difficult in laboratory
experiments to determine what parameters of the inter-particle potential effect the value of the yield stress. Here we use
the versatility of numerical simulations to study the dependence of the yield stress on the parameters of the inter-particle
potential. We find a very simple dependence on the fundamental scales which characterize the repulsive and attractive
parts of the potential respectively, and offer a scaling theory that collapses the data for widely different potentials and
in different space dimensions.
\end{abstract}
\maketitle

\begin{figure}
\centering
\hskip -1.5 cm
\includegraphics[scale = 0.70]{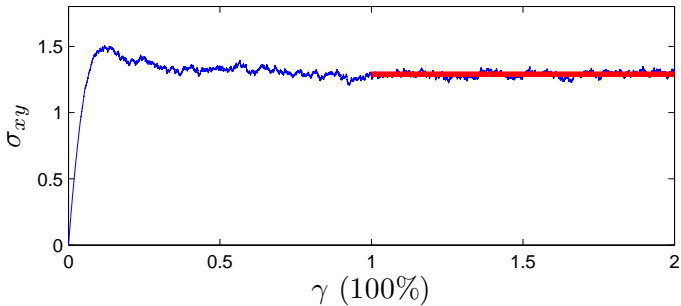}
\caption{Color online: a typical stress vs strain curve. At low strains the stress increases linearly with the slope being the
shear modulus. At a strain value of the order of 10\% the system undergoes a yielding transition \cite{10KLP}.
$\sigma_Y$ is the mean stress computed at the elasto-plastic steady state obtained after, say, 100\% strain (see red curve).}
\label{stress-strain}
\end{figure}

{\bf Introduction} At small external strain every solid reacts elastically. We are interested here in the response of amorphous
solids to high external strains, and in particular in the yield stress $\sigma_Y$ which cannot be exceeded without effecting a
plastic response which typically leads to mechanical failure via plastic flow, shear banding or fracture \cite{07SHR,04VBB,09RSa}.
In this Letter we focus on the fundamental microscopic features which determine the yield stress, using to great advantage the
versatility of numerical simulations in which the inter-particles potential can be varied at will. We work at zero temperature and
quasi-static external straining conditions, (the so-called athermal quasi-static or AQS limit) where very precise simulation results
can be obtained. To introduce the issue examine Fig. \ref{stress-strain} which exhibits a typical stress vs. strain curve, here
for a system of $N=500$ particles in AQS conditions. The yield stress is the average stress in the elasto-plastic steady state
that is obtained, say, after 100\% deformation.

Clearly, the connection of the yield stress to the fundamental microscopic characteristics of the material is of great interest,
but in fact far from being determined. To a large extent the determination of the yield stress of various amorphous solids
is still a question of trial and error without much theoretical input. From the point of view of fundamental theory we have results
for the dependence of the yield stress on the density $\tilde \rho$ only in the case of very simple inter-particle potentials which are purely repulsive, without an attractive tail. Simple results are available for potentials of the form $\phi(r)\sim r^{-\alpha}$. Defining then the re-scaled (dimensionless) density as
$\rho=\lambda^d \tilde \rho$ (where $\lambda$ is a typical length scale, cf. Eq. (\ref{potential})),
then the yield stress for solids at different dimensionless densities $\rho$ was shown to vary like
\begin{equation}
\sigma_Y \approx \frac{\varepsilon}{\lambda^d} \rho^\nu  \ , \quad \nu=1+\alpha/d \label{denscale}
\end{equation}

It had been shown \cite{09LPb} that this scaling law is not valid in general when the inter-particle potential contains an attractive part,
except at very high densities when the particles are squeezed against the repulsive part of the potential. The aim of this Letter
is to provide a scaling theory that remains valid for generic potentials with an attractive and a repulsive part.

{\bf System and potentials}: We investigate a binary system where the amount of bi-dispersity of `small' and `large' particles was chosen
to guarantee that the models produce good glass formers both in 2 and 3 dimensions. We constructed a template for inter-particle
potentials where the microscopic lengths can easily be tuned:

\begin{equation}\label{potential}
\phi\!\left(\!\sFrac{r_{ij}}{\lambda_{ij}}\!\right) \!\!=\!\!
\!\left\{
\begin{array}{ccl}
&&\!\!\!\!\!\!\!\!\!\!4\varepsilon\left[\left(\frac{\lambda_{ij}}{ r_{ij}}\right)^{12}-\left(\frac{\lambda_{ij}}{r_{ij}}\right)^{6}\right]
   \quad\quad\quad\quad\quad\ \ , \ \ \frac{r_{ij}}{\lambda_{ij}}\le \frac{r_{\rm min}}{\lambda}\\

\!\!\!\!\!\!\!\!&&\!\!\!\!\!\!\!\!\!\! \varepsilon\left[a\left(\frac{\lambda_{ij}}{ r_{ij}}\right)^{12}\!\!\!\!\!\!
                                                        -\!\!\!b \!\left(\frac{\lambda_{ij}}{ r_{ij}}\right)^{6}\!\!\!
                                                        +\!\! \displaystyle{\sum_{\ell=0}^{n}}c_{2\ell}\left(\sFrac{r_{ij}}{\lambda_{ij}}\right)^{2\ell}\right]\!\!
   , \frac{r_{\rm min}}{\lambda}\!\!<\!\!\frac{r_{ij}}{\lambda_{ij}}\!\!<\!\!\frac{r_{\rm co}}{\lambda} \\

\!\!\!\!&&\quad\quad\quad\quad 0 \ \quad\quad\quad\quad\quad\quad\quad\quad\quad
   , \quad \frac{r_{ij}}{\lambda_{ij}} \!\ge\! \frac{r_{\rm co}}{\lambda}

\end{array}
\right.\!\!
\end{equation}

where $r_{\rm min}/\lambda_{ij}$ is the length where the potential attain it's minimum, and $r_{\rm co}/\lambda_{ij}$
is the cut-off length for which the potential vanishes. The coefficients $a,~b$ and $c_{2\ell}$ are chosen such
that the repulsive and attractive parts of the potential are continuous with two derivatives at the potential minimum and the potential goes
to zero continuously at $r_{\rm co}/\lambda_{ij}$ with two continuous derivatives as well. To satisfy the latter constraints it suffices for $n$ to be
equal to 3. In other cases where $r_{\rm co}$ is fixed and we want to control the shape of the attractive part of the
potential, $n=4$.
The interaction length-scale $\lambda_{ij}$ between any two particles $i$ and $j$ is $\lambda_{ij} = 1.0\lambda$, $\lambda_{ij} = 1.18\lambda$
and $\lambda_{ij} = 1.4\lambda$ for two `small' particles, one `large' and one `small' particle and two `large' particle respectively. The
unit of length $\lambda$ is set to be the interaction length scale of two small particles, $\varepsilon$ is the unit of energy and $k_B = 1$.

\begin{figure}
\centering
\hskip -1. cm
\includegraphics[scale = 0.60]{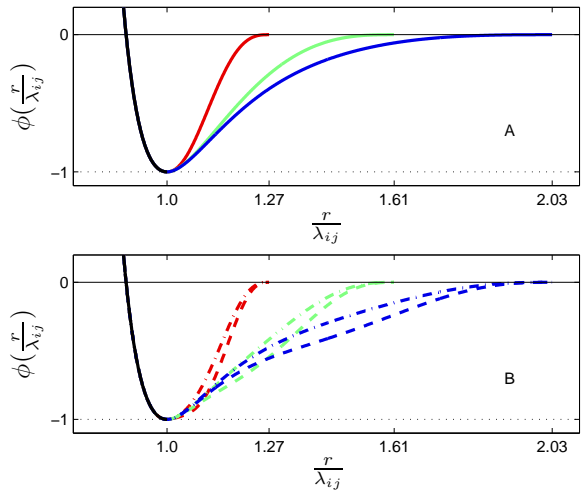}
\caption{Color online: a selection of different potentials used to determine the yield stress. Panel A depicts potentials where $r_{\rm co}/\lambda_{ij}$ is the changing length. Panel B shows potentials for which $r_{\rm co}/\lambda_{ij}$ remains unchanged but have a different attractive behavior.}
\label{potentials}
\end{figure}

{\bf Methods}: The work presented here investigates systems under simple shear in the athermal quasi-static (AQS) limit,
$T \to 0$ and $\dot\gamma\to 0$, where $\dot\gamma$ is the strain rate. AQS methods have been extensively
used recently \cite{04ML,06TLB,06MLa,07BSLJ,09LPa,08TTLB} as a tool for investigating plasticity in amorphous systems. The order in which the
limits $T \to 0$, $\dot\gamma\to 0$ are taken is important, since one expects that at any finite temperature the stress in the
system can thermally relax given long enough time \cite{08EP} (or small enough strain rates), hence the limit $T \to 0$ should be
taken prior to the $\dot\gamma\to 0$ limit. According to AQS methods, starting from a completely quenched configuration of the
system, we apply an affine simple shear transformation to each particle i in our shear cell, according to
\begin{eqnarray}
r_{ix} &\to& r_{ix} + r_{iy}\delta\gamma , \nonumber \\
r_{iy} &\to& r_{iy} ,  \label{affineTransformation}
\end{eqnarray}
in addition to imposing Lees-Edwards boundary conditions \cite{91AT}. The strain increment $\delta\gamma$ plays
a role analogous to the integration step in standard MD simulations. We choose for the discussed systems
$\delta\gamma = 5\times 10^{-5}$, which is sufficiently small for the analysis of the steady state mean values. The affine
transformation Eq. (\ref{affineTransformation}) is then followed by the minimization \cite{MIN} of the potential energy under the constraints
imposed by the strain increment and the periodic boundary conditions.

\begin{figure}
\hskip -0.0 cm
\includegraphics[scale = 0.50]{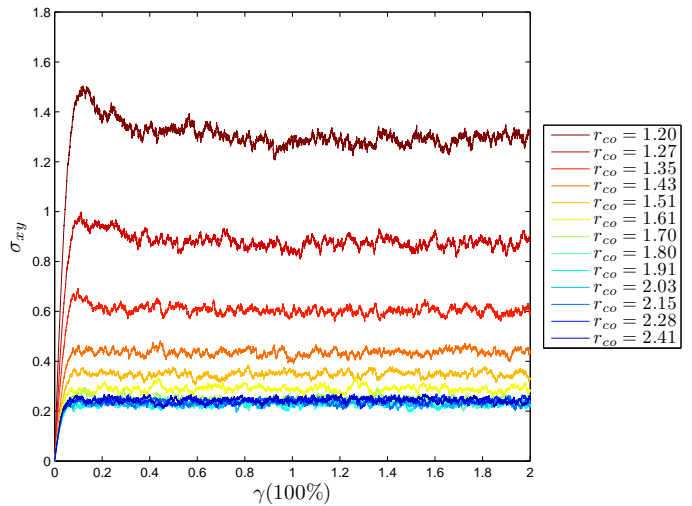}
\caption{Color online: stress vs. strain curves in 2-dimensional systems with the potential shown in Fig. \ref{potentials}, panel A. Note the convergence with increasing $r_{\rm co}$. }
\label{rawdata}
\end{figure}

{\bf The yield stress in 2-dimensions}: We prepared systems with density $\rho$ ranging in the interval $[0.775, 0.935]$ with increments of $0.02$.
The low boundary of this interval is determined by retaining positive pressure in our simulation box for any of the considered microscopic lengths.
For lower densities the system becomes porous with patches of vacuum, and we do not investigate such states. The range of the microscopic lengths
considered to generate the various potentials is
\begin{equation}
r_{\rm co} =  1.2  \lambda_{ij} \times 1.06^k\quad {\rm where} \quad k  \in \{0, ... , 12\}. \label{lengths}
\end{equation}
The microscopic lengths used to generate our data-set of $\sigma_Y$ values include all combinations of $r_{\rm co}$ and $\rho$ with $n=3$.
A few examples of the different potentials can be seen in Fig. \ref{potentials}, panel A.
Each such system with $N=500$ particles was strained in AQS conditions and a typical result for stress-strain curves with given $\rho$ is shown in Fig. \ref{rawdata}.
The yield stress in simulations are read from the steady-state stress-strain curve as depicted in Fig. \ref{stress-strain}.
To understand these $\sigma_Y$ values, we note that for a finite cut-off length the density determines how many neighboring particles are within the interaction
length, and $\sigma_Y$ is expected to be a strong function of the density. In the limit $r_{\rm co}\to \infty$ all the particles are within the
interaction range and the dependence on the density disappears in favor of a limiting value of $\sigma_Y\to\tilde \sigma_Y$.

To collapse the data on a single graph we note that there are three independent lengths in this problem, namely $\lambda$,
$r_{\rm co}$ and $\tilde \rho^{-1/d}$. These three lengths render two ratios, i.e. $s_1\equiv \lambda/\tilde \rho^{-1/d}\equiv \rho^{1/d}$ and
$s_2\equiv r_{\rm co}/\tilde \rho^{-1/d}$. In general one expects a scaling law with these two variables
\begin{equation}
\sigma_Y=\frac{\varepsilon}{\lambda^d}\rho^\nu g(s_1, s_2) \ .
\end{equation}
We expect that the limit $s_1 \to \infty$ should reproduce the high density limit Eq. ({\ref{denscale}).
In contrast, we expect that $s_2\to \infty$ should lead to the converged large interaction limit $\tilde \sigma_Y$. Thus these two dimensionless
numbers work in opposite direction, and we consider the dimensionless product $s=s_1\times s_2$ in which the effects of both ratios can balance.
These consideration lead to a scaling ansatz in the form
\begin{equation}
\sigma_Y =\frac{\varepsilon}{\lambda^d} \rho^\nu f(s) \ . \label{ansatz1}
\end{equation}

Here the scaling function $f(s)$ is expected to behave according to
\begin{equation}
f(s)=\left\{
\begin{array}{ccl}
&&{\rm const} \quad{\rm for}~s\to \infty\\
&& s^{-\nu} \quad {\rm for}~s\to 0 \ .
\end{array}
\right.,
\end{equation}
Indeed, plotting $\sigma_Y/\rho^\nu$ as a function of $s$ leads to a superb data collapse, as one can see in Fig. \ref{collapse1}.

\begin{figure}
\centering
\hskip -1.0 cm
\includegraphics[scale = 0.6]{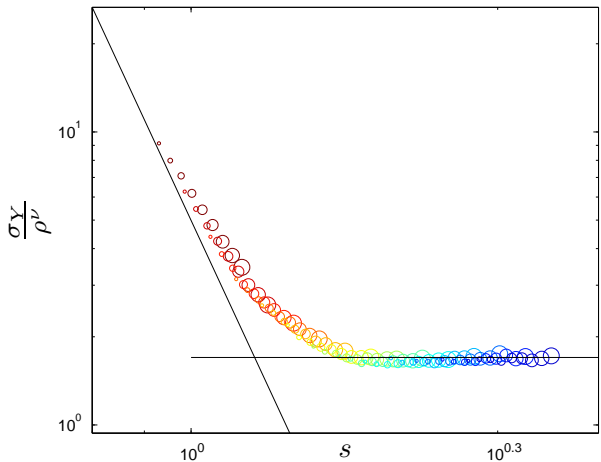}
\caption{Color online: the scaling function $f(s)$ (cf. Eq. \ref{ansatz1}), obtained by collapsing the computed $\sigma_Y$ values in 2-D. Hot colors are for shorter
values of $r_{\rm co}$, small symbols represent lower densities. Potentials used here are of the type found in Fig. \ref{potentials} panel A.}
\label{collapse1}
\end{figure}
It should be noted that the choice of the scaling variable $s=s_1\times s_2$ can be directly validated by checking that the value of
$\sigma_Y/\rho^\nu$ is indeed invariant for a given $s$ for any $s_1$ and $s_2=s/s_1$, see Fig. \ref{validation}. In the figure
we plot the scaling function $f(\tilde s=\tilde s_1\times \tilde s_2)$ vs. $f(s=s_1\times s_2)$ for a series of different values of $s$ such that $\tilde s_1\times \tilde s_2\approx s_1\times s_2$.

\begin{figure}
\centering
\hskip -1.0 cm
\includegraphics[scale = 0.6]{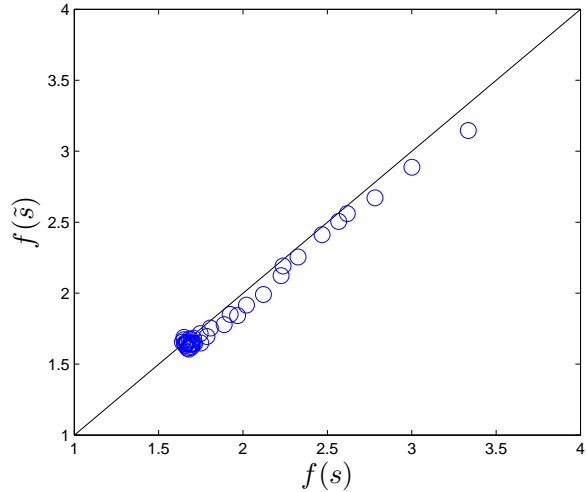}
\caption{Color online: plotted are the values of $f(s)$ vs. $f(\hat s)$ from the dataset presented in Fig. \ref{collapse1}, where $\frac{\hat s}{s}\approx 1$.}
\label{validation}
\end{figure}

{\bf The yield stress in 3-dimensions}: A sensitive test of the above mentioned scaling theory is provided by validating the explicit dimensionality dependence. To this end similar data was collected for samples in 3-dimensions using the potentials with $n=3$ and varying $r_{\rm co}$ values, see Eq. (\ref{lengths}), combined with densities in the interval
$[0.76, 1]$ with increments of $0.03$ (emanating from the above-mentioned considerations). The beautiful scaling collapse in Fig. \ref{collapse2}, simply changing the numerical value of $\nu$ as dictated by Eq. (\ref{denscale}) confirms the choice of scaling function, Eq. (\ref{ansatz1}).

\begin{figure}
\centering
\hskip -1.0 cm
\includegraphics[scale = 0.6]{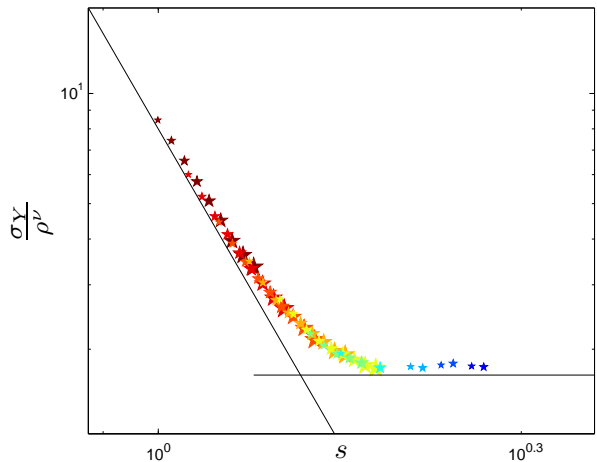}
\caption{Color online: the scaling function $f(s)$ (cf. Eq. \ref{ansatz1}), obtained by collapsing the computed $\sigma_Y$ values in 3-dimensions using potentials as in
Fig. \ref{potentials} panel A and N=1000. Hot colors are for shorter values of $r_{\rm co}$, small symbols represent lower densities.}
\label{collapse2}
\end{figure}

{\bf The effect of potential shape on the yield stress in 2-dimensions}: We have so far been successful in identifying the crucial microscopic
lengths involved in determining the yield stress. The role of a change of shape in the attractive part of the potential now remains to be elucidated.
(However, regardless of the shape of the potential we expect the limiting values to remain unchanged. In other words, we still expect that the limit
$s_1 \to \infty$ should reproduce the high density limit and that $s_2\to \infty$ should lead to the converged large interaction limit $\tilde \sigma_Y$.)
To this end we used two variations of potentials with $n=4$ and $r_{\rm co}$ as in Eq. (\ref{lengths}). For examples of such potentials see Fig. \ref{potentials}, panel B.

The change in physics as a result of changing the shape of the attractive part of the potential can be easily understood. As the potential well widens, the slope of the potential on the way to $r_{\rm co}$ increases.
As long as the second shell of particles is {\bf outside} the interaction range, the system softens as a result of the
widening well. As soon as  $r_{\rm co}$ reaches a value that {\bf includes} the second shell, the strong attractive forces stiffen the system, thus creating a dip in the scaling function.

To corroborate this picture we repeated the AQS straining experiments for the  altered potentials (see lower panel of Fig. \ref{potentials} at similar densities as before in 2-dimensions. In Fig. \ref{collapse3} we see the unchanging asymptotes of the scaling function and the expected dip in $\sigma_Y/\rho^\nu$. Note the larger dip in the
right panel of Fig. \ref{collapse3}. Note that this effect of softening and then hardening is always
present even if it is barely noticeable \cite{footnote1}.

\begin{figure}[h]
\centering
\hskip -1.0 cm
\includegraphics[scale = 0.45]{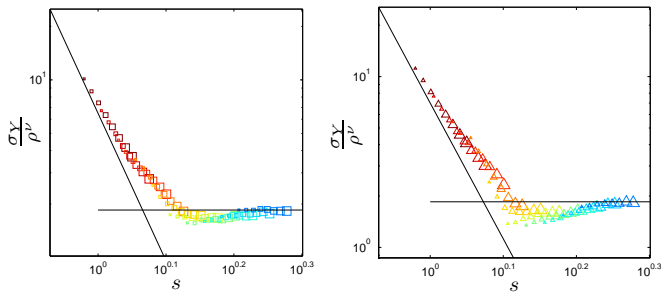}
\caption{Color online: the scaling function $f(s)$ (cf. Eq. \ref{ansatz1}), obtained by collapsing the computed $\sigma_Y$ values in 2-D. Hot colors are for shorter
values of $r_{\rm co}$, small symbols represent lower densities. Left panel is the collapse of $\sigma_Y$ values computed using the `dash-dot' fashion potentials
appearing in Fig. \ref{potentials} panel B and the right panel is the collapse of $\sigma_Y$ values computed using the `dashed' fashion potentials in the same panel.}
\label{collapse3}
\end{figure}

{\bf Concluding remarks}: We have presented a scaling theory of the yield stress in AQS conditions, stressing the data
collapse, and therefore of the high predictability of the scaling theory, when the potentials are changing widely both in
their shape and in the cutoff scale. The data collapse means that measuring the data for one potential we can predict
how the yield stress will change for other potentials. Together with the recent scaling theory for the flow stress as
a function of temperature and strain rate (albeit at this point only for purely repulsive potentials \cite {09LPb}), we begin to see a powerful theory based on scaling concepts that is emerging
for the discussion of plasticity in amorphous solids.

This research was supported in part by the Israel Science Foundation, The German Israeli Foundation, the IMOS under
the bilateral program with France and the European Research Council under Grant \#267093 STANPAS.


\begin{thebibliography}{99}
\bibitem{footnote1}
Note that slight increase in $\sigma_Y$ for the largest value of $r_{\rm co}$ in Fig. \ref{rawdata}. This is due to the
influence of second shell neighbors entering the interaction range.
\end{thebibliography}
\end{document}